\documentclass[prd,preprint,floatfix,amsmath,superscriptaddress,amssymb,showkeys,oldgerm,showpacs]{revtex4}
\usepackage{graphicx,dcolumn,bm}
\begin{document}
\title{ Bergmann-Thomson energy-momentum complex for  solutions more general 
than the Kerr-Schild class}
\author{ S.~S.~Xulu}
\affiliation{Department of  Computer Science, University of Zululand, 
Private Bag X1001, 3886 Kwa-Dlangezwa, South Africa}
\date{\today}

\begin{abstract}
In a very  well-known  paper, Virbhadra's research group proved that  the  
Weinberg,
Papapetrou, Landau and Lifshitz, and   Einstein energy-momentum complexes 
``coincide'' for all metrics of Kerr-Schild class.
A few years later, Virbhadra clarified that this ``coincidence'' in fact 
holds  for  metrics more
general than the Kerr-Schild class.  In the present paper, this  study is 
extended  for the
Bergmann-Thomson complex and it is  proved that this  complex also 
``coincides'' with those complexes  for
a more general than the Kerr-Schild  class metric.
\end{abstract}



\maketitle

\section{Introduction}
Virbhadra and his collaborators (notably, Professor Nathan Rosen of the EPR 
paradox,  Einstein-Rosen bridge, and Einstein-Rosen gravitational waves 
fame)
resurrected the subject of energy-momentum in general 
relativity\cite{VirColl,RosVir,ChamVir,Vir}.
They proved that energy-momentum complexes ``coincide''
and give reasonable results for a particular space-time. They found this to 
be true for some well-known
and physically significant  space-times.  Further, Aguirregabiria, Chamorro 
and Virbhadra\cite{Agu96}
observed  that the energy-momentum
complexes of Weinberg, Papapetrou, Landau and Lifshitz (LL), and Einstein 
all ``coincide'' for any metric of Kerr-Schild class if calculations are 
accomplished  in Kerr-Schild Cartesian coordinates. A few
years later,  Virbhadra\cite{Vir99} noted
that this ``coincidence '' is in fact true for space-times more general than 
the Kerr-Schild class.
This  paper by Virbhadra  triggered a lot of  fascinating publications in 
this subject in international
journals and the  list is growing   exponentially. It is really impossible 
to discuss most of them in this
paper. In the following paragraph, we are able to narrate only some of the 
results.

We studied the energy of the  Kerr-Newman metric, Bianchi Type I universes, 
Schwarzschild black hole
in a magnetic universe, and nonstatic spherically symmetric metrics\cite{X}. 
Vagenas
investigated the energy of a radiating charged particle in Einstein's 
universe and  a dyadosphere
of a Reissner-Nordstr\"{o}m  black hole\cite{vag1}. He also studied the 
energy distribution in (2+1)
dimensional space-times\cite{vag2}. Gad\cite{Gad} obtained the  energy 
density associated with solutions
exhibiting directional singularities. Sharif\cite{Sharif} computed  the 
energy distributions for  a
regular black hole space-time, the G\"odel universe, and the Weyl metric. 
Vargas\cite{vargas} calculated the
energy of the universe in  tele-parallel gravity.

Aydogdu and his colleagues studied energy of the universe in Bianchi type-I 
and II  models,
Reboucas-Tiomno-Korotkii-Obukhov, Godel, and some other 
space-times\cite{Aydogdu}.
Aygun and his collaborators computed the energy and momentum distributions 
in Szekeres type I
and II space-times and the  Marder space-time\cite{Aygun}. 
Halpern\cite{Halp} obtained the energy associated with
black plane and Taub cosmological solutions. Further, Salti and his 
collaborators have accomplished gigantic amount of
work in this field of research\cite{Salti}. They studied energy and momentum 
problem for the Schwarzschild de Sitter
space-time, Reissner-Nordstr\"om anti-de Sitter black holes, closed 
universes, a charged wormhole,  and  for several other
interesting cases. For a comprehensive review on the energy-momentum problem 
in general relativity and some recent
important  results on this topic, see the Ph.D. thesis of the present 
author\cite{MyThesis}.

We discussed in the first paragraph of this section that  
Virbhadra\cite{Vir99} showed that the energy-momentum complexes of Weinberg, 
  Papapetrou, Landau and Lifshitz,
and Einstein  ``coincide'' for a more general metric than the Kerr-Schild
class, subject to the condition that  calculations are performed in 
Kerr-Schild Cartesian coordinates. In a  recent paper published in 
Foundation
of Physics Letters, we\cite{X06} proved that the Bergmann-Thomson complex 
furnishes the same result for the Kerr-Newman black hole
metric as obtained by Aguirregabiria et al.\cite{Agu96} about a decade ago. 
As a natural flow of research curiosity, results
obtained by Virbhadra\cite{Vir99} now   tempts
us to investigate whether or not the Bergmann-Thomson complex also yields 
the same result for metrics more general than the Kerr-Schild
class or for at least  the Kerr-Schild class. As in our previous papers, we 
use geometrized units and follow the convention that Latin and Greek indices
respectively take values 0 to 3  and 1 to 3.

\section{Energy-momentum complexes of  Weinberg,
Papapetrou, Landau and Lifshitz, and   Einstein}

A renowned particle  physicist and Nobel laureate Steven Weinberg proposed 
an energy-momentum
complex (see in \cite{MyThesis}). This is now termed    the {\em Weinberg  
energy-momentum complex} and  is expressed
by the equation

\begin{equation}
{\cal W}^{ik}= \left(\frac{1}{16\pi}\right) {\omega^{qik}}_{,q} \text{.}
\end{equation}

In the above equation,

\begin{equation}
\omega^{qik}= \frac{\partial h^r_{\ r}}{\partial x_q}\eta^{ik}
               + \frac{\partial h^{ri}}{\partial x^r}\eta^{qk}
         + \frac{\partial h^{qk}}{\partial x_i}
          -  \frac{\partial h^r_{\ r}}{\partial x_i}\eta^{qk}
         - \frac{\partial h^{rq}}{\partial x^r}\eta^{ik}
         - \frac{\partial h^{ik}}{\partial x_q}
\end{equation}

where
\begin{equation}
h_{ik}=g_{ik} - \eta_{ik}.
\end{equation}

It  must be remembered that the  indices on  $\partial/
\partial x_i$ and  $h_{ik}$  are lowered/raised   with the aid of $\eta$'s,
where $\eta_{ik}$ is the Minkowski metric in $3+1$ space-time dimensions 
such that
$\eta^{ab} = diag(1,-1,-1,-1)$.

Equation (2) shows that the pseudotensor $\omega^{qik}$ is antisymmetric in 
its first two
indices, i.e.

\begin{equation}
\omega^{qik}=- \omega^{iqk}.
\end{equation}

This antisymmetry property helps  applying Gauss's theorem and computing   
energy and momentum  inside a closed
surface. The Weinberg
complex  ${\cal W}^{ik}$ is symmetric.
The energy and momentum components  can be computed using the following 
formula:
\begin{equation}
{\mathbb P}^k=\int\!\!\!\int\!\!\!\int{ {\cal W}^{k0}\, dx^1 \, dx^2 \, 
dx^3} .
\end{equation}
If we use Gauss's theorem in the above equation, we get
\begin{equation}
{\mathbb P}^k=\left(\frac{1}{16 \pi}\right) \int\!\!\!\int{ \omega^{\gamma 
0k} \  n_{\gamma}\   d\Omega} .
\end{equation}

$d\Omega$ is the infinitesimal surface element and  $n_{\gamma}$ is the unit 
normal vector to that surface.
In equation (5), ${\cal W}^{00}$ and ${\cal W}^{\alpha 0}$   are 
respectively the energy and
momentum  density components. It was indicated in the Introduction that the 
Greek indices run from 1 to 3.

A leading Greek  relativist A. Papapetrou also obtained a symmetric 
energy-momentum complex,
now known as the {\em  Papapetrou complex} (see in \cite{MyThesis}). This is 
given by the following expression:

\begin{equation}
{\mathfrak{P}}^{ik}=\left(\frac{1}{16 \pi}\right){{\cal{A}}^{ikqr}}_{,qr}.
\end{equation}
${\cal A}^{ikqr}$ in the above equation is
\begin{equation}
{\cal A}^{ikqr}=\sqrt{-g} \left(g^{ik} \eta^{qr}  + g^{qr} \eta^{ik}- g^{iq} 
\eta^{kr} - g^{qk} \eta^{ir}\right),
\end{equation}
where

\begin{equation}
\eta^{ij} =
\left(
\begin{array}{cccc}
1 & 0 & 0 & 0 \\
0 & -1 & 0 & 0 \\
0 & 0 & -1 & 0 \\
0 & 0 & 0 & -1
\end{array}
\right).
\end{equation}

${\mathfrak{P}}^{00}$ and ${\mathfrak{P}}^{\alpha 0}$  represent  the energy 
  and energy
current  density components.

If the metric under study is  time-independent, then we have an advantage 
that we can apply
Gauss's theorem  to calculate energy and momentum components.
\begin{equation}
{\mathbb P}^k= \left(\frac{1}{16 \pi}  \right)
\int\!\!\!\int{{{\mathcal A}^{k0\gamma\beta}}_{,\beta} \  n_{\gamma}\  
d\Omega} .
\end{equation}

One of the greatest Russian theoretical physicist and Nobel laureate Landau 
and his collaborator Lifshitz discovered
a symmetric energy-momentum complex (see in \cite{MyThesis}). This is called 
the {\em Landau and Lifshitz energy-momentum complex} and
is given by the following  cute mathematical expression:
\begin{equation}
{\cal L}^{ik}=  \left(\frac{1}{16 \pi}\right) {{\Lambda}^{ikqr}}_{,qr},
\end{equation}
with
\begin{equation}
{\Lambda}^{ikqr}=-g \left(g^{ik} g^{qr}-g^{iq} g^{kr}\right).
\end{equation}
${\cal L}^{00}$ and $ {\cal L}^{\alpha 0}$  furnish  the energy and
momentum density components associated with a  given metric.
The energy and momentum  are
\begin{equation}
{\mathbb P}^k=\int\!\!\!\int\!\!\!\int{{\cal L}^{k0}\,dx^1\,dx^2\,dx^3} .
\end{equation}
A straightforward application of  Gauss's theorem  to the integral on the 
right hand side of the above equation yields :
\begin{equation}
{\mathbb P}^k=  \left(\frac{1}{16 \pi}\right)
   \int\!\!\!\int{ {{\Lambda}^{k0\gamma q}}_{,q} \  n_{\gamma}\ d\Omega} .
\end{equation}

We  now discuss Einstein's complex. Einstein was in fact  the first to 
obtain an
energy-momentum complex. However, this complex was not symmetric like 
aforesaid three we discussed
and therefore cannot  be used for defining angular momentum in general 
relativity. This complex is
\begin{equation}
{\cal E}_i{}^{k} = \left(\frac{1}{16 \pi}\right)  {\cal H}^{\ kq}_{i \ \ 
,q}.
\end{equation}
${\cal H}_i^{\ kl}$ in the above equation is
\begin{equation}
{\cal H}_i^{\ kq}\  =\ - {\cal H}_i^{\ qk}\ =\  \frac{g_{is}}{\sqrt{-g}}
         \left[-g \left( g^{ks} g^{qr} - g^{qs} g^{kr}\right)\right]_{,r} \ 
.
\end{equation}

The energy and momentum components  are
\begin{equation}
{\mathbb P}_k \ = \ \int \int \int {\cal E}_k^{\ 0} dx^1  dx^2 dx^3 .
\end{equation}
This equation gives
\begin{equation}
{\mathbb P}_k\ =\ \left(\frac{1}{16 \pi}\right) \ \int\int\ {\cal H}_k^{\ 0 
\gamma} \ n_{\gamma}\ d\Omega .
\end{equation}

\section{Kerr-Schild class space-times }

Let us denote a scalar field by ${\cal S}$ and a vector field by $V_i$. Then 
a Kerr-Schild class
metric is well-known to be defined by the metric  $g_{ab}$ as follows:
\begin{equation}
g_{ab} = \eta_{ab} - {\cal S} \ V_a V_b ,
\end{equation}
where $\eta_{ab}$ is the Minkowski metric as defined in equation $(9)$ and 
the vector field $V_a$ satisfies the following conditions
in the Minkowski space-time:
\begin{eqnarray}
      \eta^{qr} V_q V_r &=& 0 ,  \text{\hspace*{1.0in} (null condition)}  
\nonumber\\
       \eta^{qr} V_{a,q} V_r &=& 0 ,\text{\hspace*{1.0in} (geodesic  
condition)} \nonumber \\
       \left(V_{q,r} + V_{r,q}\right) {V^q}_{,s} \  \eta^{rs}
- \left({V^q}_{,q}\right)^2 &=& 0 \text{\hspace*{1.0in} (shear-free 
condition)}
\end{eqnarray}
Virbhadra\cite{Vir99} mentioned a few  important examples of the Kerr-Schild 
class space-times:\\
\hspace*{0.3in}   $\bullet$   Schwarzschild        \hspace*{1.9in}     
$\bullet$   Vaidya  \\
  \hspace*{0.3in}  $\bullet$   Reissner-Nordstr\"{o}m   \hspace*{1.5in}  
$\bullet$   Bonnor-Vaidya\\
  \hspace*{0.3in}  $\bullet$   Kerr                    \hspace*{2.6in}   
$\bullet$   Vaidya-Patel\\
  \hspace*{0.3in}  $\bullet$   Kerr-Newman             \hspace*{1.9in}   
$\bullet$   Dybney {\it et al.} \\
For a  comprehensive  discussion of these space-times, see  references in 
\cite{Vir99}.
Aguirregabiria, Chamorro and Virbhadra\cite{Agu96} discovered  a  marvelous  
result that the energy-momentum
complexes  of Weinberg, Papapetrou, Landau and Lifshitz and Einstein
``coincide'' for any space-time of the Kerr-Schild class. They established  
the following
extremely important  relationship among energy-momentum complexes:
\begin{eqnarray}
{\cal W}^{ik} &=& {\mathfrak P}^{ik} = {\cal L}^{ik} =
\left(\frac{1}{16\pi}\right) {\mathbb U}^{ikrs}{}_{,rs}, \nonumber  \\
{\cal E}_i{}^{k} &=&   \eta_{iq} {\cal L}^{qk} \hspace*{.1in}\text{,}
\end{eqnarray}
with
\begin{eqnarray}
{\mathbb U}^{ikrs}   =  {\cal S}  \left(
\eta^{ik} V^r V^s-\eta^{ir} V^k V^s +\eta^{rs} V^i V^k-\eta^{ks} V^i V^r
\right).
\end{eqnarray}
These results shook the prevailing notion that different energy-momentum 
complexes
will  meaninglessly give different results for a given metric.
Using the above results, they\cite{Agu96}  further obtained general 
expressions for the energy, momentum
and angular momentum for any metric of the Kerr-Schild class.

It seems that Aguirregabiria, Chamorro and Virbhadra\cite{Agu96} did  not 
notice whether or not all
the three conditions (null, geodesic and shear-free) were used for obtaining 
the Eq. $(21)$  with $(22)$.
Thanks to Virbhadra\cite{Vir99} that three years  later he
noticed and reported that only null and geodesic
conditions (not the shear-free) were  used to derive the relationship given 
in Eq. $(21)$. Thus,
this relationship is true for space-times more general than the Kerr-Schild 
class, because the shear-free condition
is not demanded  for this derivation.

\section{Bergmann-Thomson complex  for a class of space-times more general 
than the Kerr-Schild class}

About more than fifty years ago, Bergmann and Thomson\cite{BT}  obtained a 
new energy-momentum complex:

\begin{equation}
{\cal{B}}^{jk} = \frac{1}{16\pi}  {\bf B}^{jkq}_{\quad ,q}
\text{,}
\end{equation}
where
\begin{equation}
{\cal B}^{jkq} =  g^{jr} {\cal V}_r^{\  kq}
\end{equation}
with
\begin{equation}
{\cal V}_r^{\  kq}\  =\ - {\cal V}_r^{\  qk}\ =\
\frac{g_{rs}}{\sqrt{-g}}
         \left[-g \left( g^{ks} g^{qp} - g^{qs} g^{kp}\right)\right]_{,p} \
.
\end{equation}

Similar to the four energy-momentum complexes  discussed in the last 
Section, ${\bf{B}}^{jk}$   does not
transform as a tensor under a general
coordinate transformation. ${\cal{B}}^{00}$ is the energy density and  
${\cal{B}}^{\alpha0}$
is the momentum density components.

The energy and momentum components ${\mathbb P}^k$ are given by

\begin{eqnarray}
  {\mathbb P}^k &=&  \int \int \int {\cal{B}}^{k0} dx^1 dx^2 dx^3   
\nonumber \\
                &=& \left( \frac{1}{16\pi} \right)  \int \int {\bf 
B}^{k0\gamma} n_{\gamma} d\Omega \text{.}
\end{eqnarray}

We\cite{X06} computed  energy and momentum distributions in the Kerr-Newman 
space-time  using the Bergmann-Thomson
complex and to a great wonder it came to be the  same as obtained by 
Aguirregabiria, Chamorro and Virbhadra\cite{Agu96}
in formulations of Weinberg, Papapetrou, Landau and Lifshitz, and Einstein. 
Our result attracts us to investigate further if
the  Bergmann-Thomson complex ``coincides'' with other complexes for  any 
Kerr-Schild class space-times or for a
more general class than this.

Let us examine  all complexes discussed in Section 2 and the  
Bergmann-Thomson complex  meticulously. It is straightforward to prove and
is also known  (for instance, see in \cite{Agu96}) that for the metric 
expressed by Eq. $(19)$, one has
\begin{equation}
g = - 1 .
\end{equation}
It must be remembered that none of the three conditions in   Eq. $(20)$ is 
required to obtain the above equation. Only Eq.
$(19)$ is enough to derive the  equation $(27)$.
Now equations $(11)$ and $(23)$  with the equation $(27)$ results
\begin{equation}
{\cal L}^{ik}  =  {\cal  B}^{ik} .
\end{equation}

Thus, equations $(21)$ and  $(28)$ and Virbhadra's result\cite{Vir99} that 
only null and geodesic conditions are required to derive
the Eq. $(21)$, we arrive at the following result.

For any space-time more general than the Kerr-Schild class (i.e. space-time 
described by Eq. $(19)$ satisfying only the
null and geodesic conditions of Eq. $(20)$), one gets

\begin{eqnarray}
{\cal W}^{ik} &=& {\mathfrak P}^{ik} = {\cal L}^{ik} = {\cal B}^{ik} =
\left(\frac{1}{16\pi}\right) {\mathbb U}^{ikrs}{}_{,rs}, \nonumber  \\
{\cal E}_i{}^{k} &=&   \eta_{iq} {\cal L}^{qk} \hspace*{.1in}\text{,}
\end{eqnarray}
with
\begin{eqnarray}
{\mathbb U}^{ikrs}   =  {\cal S}  \left(
\eta^{ik} V^r V^s-\eta^{ir} V^k V^s +\eta^{rs} V^i V^k-\eta^{ks} V^i V^r
\right).
\end{eqnarray}

\section{Summary}

The concept of energy-momentum distribution in general relativity has not 
been taken seriously  mainly for the following   reasons:

\begin{description}
  \item{\large \bf (a)}  Due to the liberty granted by the divergence 
relation  of an energy-momentum complex, many complexes have been proposed 
and many more can
          be discovered. There is no unique definition.
   \item{\large\bf(b)} These complexes are pseudotensors (non-tensors under 
general coordinate transformations). Therefore, an energy-momentum
          complex can in principle give different results in different 
coordinates for any space-time. For instance, M\o ller investigated the
           Schwarzschild metric in Einstein's prescription. One would be 
really shocked
          that the total energy
          diverges (i.e.  $-\infty$)  in spherical polar coordinates, 
whereas it is the expected Schwarzschild mass $M$ if
           computations are accomplished in quasi-Cartesian coordinates 
(refer to \cite{MyThesis} for more information.)
    \item{\large\bf(c)} Misner, Thorne, and Wheeler\cite{MTW}  expressed 
that to look for a local gravitational energy-momentum is looking for the 
right answer
            to the wrong question;  however, for spherical systems the 
gravitational potential energy is correct and meaningful.
\end{description}
Because of  these reasons, this subject remained  neglected  for a very long 
period of time until under the leadership of Virbhadra this
subject was re-animated. We write a few important points about the recent 
development.
\begin{description}
  \item{\large \bf (i)} Virbhadra showed for several space-times that 
different energy-momentum complexes give the same and reasonable results 
(see in \cite{VirColl}.)
   \item{\large\bf(ii)} A leading relativist {\em Bondi}\cite{Bon} expressed 
  his viewpoint: ``In relativity, a non-localizable form of energy is 
inadmissible,
         because any form of energy contributes to gravitation and so its
         location can in principle be found.''
    \item{\large\bf(iii)} A renowned   astrophysicist and  Nobel laureate 
{\em S. Chandrasekhar}  showed interest in the energy-momentum 
complexes\cite{Cha}.
   \item{\large\bf(iv)} Aguirregabiria, Chamorro and Virbhadra\cite{Agu96}  
showed that the energy-momentum complexes of Weinberg, Papapetrou, Landau
         and Lifshitz, and Einstein give the
         same results for any Kerr-Schild class metric.
    \item{\large\bf(v)} Rosen (am eminent collaborator of {\em Albert 
Einstein}) and Virbhadra and then again Virbhadra (see in \cite{RosVir} and 
\cite{VirColl}) showed that different energy momentum complexes give
       the same and reasonable results for the Einstein-Rosen space-time, 
which is though not of Kerr-Schild class.
       Clearly, the ``coincidence'' of different complexes is not confined  
to the Kerr-Schild class space-times.
    \item{\large\bf(vi)} Virbhadra\cite{Vir99} clarified that the 
``coincidence'' of  Weinberg, Papapetrou, Landau and Lifshitz, and Einstein 
is true for a class of space-times
          more general than the Kerr-Schild class.
     \item{\large\bf(vii)} The  invaluable contributions of Virbhadra's 
research team  enlivened this subject and now many researchers from 
different countries
        started working in this field.
     \item{\large\bf(viii)} In this paper, we (the present author) extended 
the work of Virbhadra\cite{Vir99} to the case of the Bergmann-Thomson 
complex.
\end{description}

Despite  these fascinating   successes, the energy-momentum distribution in 
a curved space-time is
far from  settled and much more painstaking efforts are
still warranted. This continues to be a  very  `hot'  topic of research in 
Einstein's general relativity.

\acknowledgments I am indeed very  grateful to my  colleagues for  their 
meticulous
reading of the manuscript and suggesting  important corrections and
modifications. I am also  thankful  to  NRF of my country (South Africa)  
for financial help without that this
work  was not possible.

\end{document}